\documentclass[pra,twocolumn,showpacs,amsfonts,amsmath,amssymb,superscriptaddress]{revtex4}

\usepackage{graphicx}
\usepackage{amsfonts}
\usepackage[latin1]{inputenc} 

\usepackage{color}

\begin{document}

\title{Angular-resolved photon-coincidence measurements in a multiple-scattering medium}

\author{Stephan Smolka}
\email{smolka@phys.ethz.ch}
\thanks{Present address: Institute of Quantum Electronics, ETH Z\"urich, 8093 Z\"urich, Switzerland}
\affiliation{DTU Fotonik, Department of Photonics Engineering, Technical University of Denmark, {\O}rsteds Plads 343, 2800 Kgs. Lyngby, Denmark}

\author{Otto L. Muskens}
\affiliation{FOM Institute for Atomic and Molecular Physics, Science Park 104, 1098 XG Amsterdam, The Netherlands}
\affiliation{SEPnet and the Department of Physics and Astronomy, University of Southampton, Southampton, SO17 1BJ, UK}

\author{Ad Lagendijk}
\affiliation{FOM Institute for Atomic and Molecular Physics, Science Park 104, 1098 XG Amsterdam, The Netherlands}

\author{Peter Lodahl}
\email{pelo@fotonik.dtu.dk}
\homepage{\\ http://www.fotonik.dtu.dk/quantumphotonics}
\affiliation{DTU Fotonik, Department of Photonics Engineering, Technical University of Denmark, {\O}rsteds Plads 343, 2800 Kgs. Lyngby, Denmark}

\begin{abstract}
We present angular-resolved correlation measurements between photons after propagation through a three-dimensional disordered medium. The multiple scattering process induces photon correlations that are directly measured for light sources with different photon statistics. We find that multiple scattered photons between different angular directions with angles much larger than the average speckle width are strongly correlated. The time dependence of the angular photon correlation function is investigated and the coherence time of the light source is determined. Our results are found to be in excellent agreement with the continuous mode quantum theory of multiple scattering of light. The presented experimental technique is essential in order to study quantum phenomena in multiple scattering random media such as quantum interference and quantum entanglement of photons.
\end{abstract}


\maketitle

When coherent light propagates through a randomly disordered medium such as biological tissues, turbid media or powders, the light waves are multiple scattered and interfere~\cite{physreports238p135,revmodphys71p313}. The hallmark of multiple scattering is the formation of a complex intensity speckle pattern in the light transmitted and reflected through the medium. Although these speckles exhibit large intensity fluctuations and appear to be random in space and frequency, these fluctuations can be statistically correlated. The resultant intensity correlations provide important information on the wave transport parameters such as the diffusion constant of light~\cite{prl61p834}. In the diffusive regime of multiple scattering, the light transport can be described by a diffusion equation and interference effects vanish after ensemble averaging over all realizations of disorder. The ensemble average can be experimentally realized by studying many independent speckle patterns corresponding to different sample configurations. However, in very strongly multiple-scattering media, deviations from diffusive light transport appear leading to non-vanishing mesoscopic intensity correlations and ultimately to Anderson localization. These effects can be explained by classical wave interference, neglecting the quantum nature of light.

An intriguing question both for reasons of fundamental interest and potential applications is how multiple scattering affects the quantum properties of light. Triggered by the theoretical quantum optical framework of multiple scattering~\cite{prl81p1829}, many proposals have been put forward addressing new phenomena that are of purely quantum origin. In particular, the contributions from quantum vacuum fluctuations have to be considered in a quantum optical description. It has been proposed that spatial quantum correlations can be induced by multiple scattering~\cite{prl95p173901} arising between two separated spatial or angular optical modes in the far field of the multiple scattered light. By extending this theory to several incident quantum states, it was predicted that quantum interference may be induced by multiple scattering~\cite{prl102p193601,prl105p163905,arXiv10072085v2} and even survive the process of ensemble averaging~\cite{prl105p090501}, which is of potential interest for quantum information processing. Recent theoretical work demonstrated the use of multiple scattering to enhance the information capacity for optical communication by exploring the intrinsic coupling between input modes and output modes~\cite{science287p287}. This effect can be even further increased exploiting the quantum optical aspects of multiple scattering using nonclassical light sources that exhibit sub-Poissonian photon statistics~\cite{prl89p043902,prl102p193901}.

The quantum optical properties of multiple scattered photons can be accessed experimentally through their statistics and correlations~\cite{prl94p153905,prl97p103901,prl102p193901,prl104p173601,science327p1352}. So far, only initial experimental studies were carried out in the quantum regime of multiple scattering using nonclassical light sources: spatial quantum correlations in the multiply scattered light were observed using squeezed light. The correlation function was determined by analyzing total transmission quantum noise measurements resulting into an average over all spatial and angular directions~\cite{prl102p193901}. Recently, Peeters \emph{et al}. observed two-photon speckle patterns by sending two spatially entangled photons through diffusers~\cite{prl104p173601} and spatial correlations were observed in photon coincidence measurements exploiting surface scatterer for a single configuration of disorder.

Here, we access for the first time photon correlations of multiple scattered light in three-dimensional random media directly using angular-resolved photon coincidence measurements between separate multiple scattering channels. The angular photon correlation function is induced by multiple scattering of light and constitutes the multichannel analogue of the Hanbury Brown and Twiss experiment generalized to a multiple scattering setting~\cite{nature177p27}. So far, such an extension of the two-channel correlation function has been actively studied in the realm of electron transport in mesoscopic conductors~\cite{physreports336p1}. We use a pseudothermal light source to investigate the angular dependence and record the angle and time correlations of the multiply scattered photons. After a configurational average over realizations of disorder, we find that the angular photon correlation function is dependent on the photon statistics of the light source and on classical intensity correlations induced by the multiple scattering medium. Our experimental results demonstrate the validity of the general quantum formalism of multiple scattering based on a continuous mode theory in the time domain. We underline that our measurement scheme is not only restricted to the regime where a classical description of multiple scattering suffices but could pave the way to study quantum interference phenomena, as theoretically proposed in Ref.~\onlinecite{prl105p090501}. Furthermore, the investigation of photon correlations in speckle patterns could be of importance for the field of ghost imaging~\cite{physreva52p3429,prl94p063601,prl94p183602}.

\begin{figure}[]
  \centering
  \includegraphics[width=0.45\textwidth]{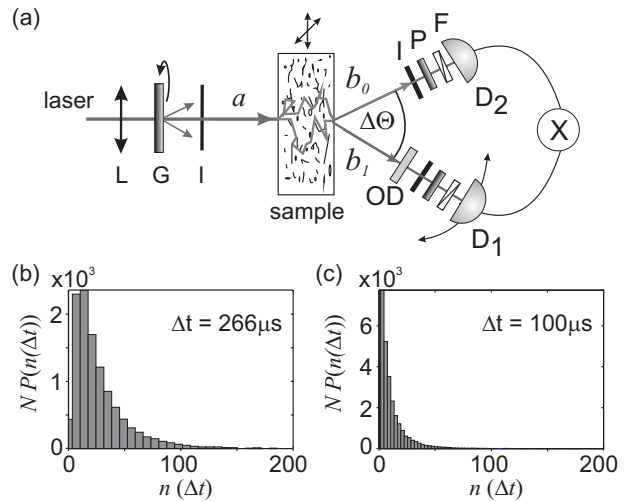}
\caption{\textbf{a}, Sketch of the experimental setup. L: lens, G: rotating ground glass plate, I: iris, P: polarizer, F: $10\:\text{nm}$ interference filter, $\Delta\Theta$: angle between detector D$_1$ and D$_2$; $a$: incident light channel; $b_0$, $b_1$: exit light channels. Optionally a neutral density filter (OD) is inserted in front of detector D$_1$ attenuating the light beam. The relative angle between D$_2$ and the optical axis of the incident light beam is 4.5$^\circ$. To probe different realizations of disorder the sample is displaced. \textbf{b}, Recorded photon counting distribution, $P(n(\Delta t))$, of the light source on detector D$_1$ after the neutral density filter and without the sample using a time interval of $\Delta t=266\,\mu$s to bin the measured photons. The photon counts $n(\Delta t)$ have been measured $N = 11.000$ times. \textbf{c}, Same measurement as in \textbf{b}, but using a time interval of $\Delta t=100\,\mu$s.}
  \label{fig01}
\end{figure}

The dimensionless angular photon correlation function between two angular directions $b_0$ and $b_1$ (see Fig~\ref{fig01}), is defined as
\begin{equation} \overline{C_{b_0b_1}^{Q}(t,t',\Delta t)} = \frac{\overline{\left<:\hat{n}_{b_0}(t,\Delta t)
\hat{n}_{b_1}(t',\Delta t):\right>}}{\overline{\left<\hat{n}_{b_0}(t,\Delta t)\right>}
\times \overline{ \left<\hat{n}_{b_1}(t',\Delta t)\right>}}, \label{eq01}
\end{equation}
where the classical ensemble average of the stochastic process of multiple scattering is denoted by the bars. Performing the ensemble average enables extracting statistical correlations that are hidden in the fluctuations of the speckle pattern. The normal ordered correlation function in the numerator can be measured by recording the coincidence counts between two detectors within the time interval $\Delta t$~\cite{scully}. The dimensionless mean photon number is obtained by integrating the photon flux operator over the measurement time $\Delta t$:   $\hat n(t,\Delta t) = \int_t^{t+\Delta t}dt'\,\hat a^\dag(t')\hat a(t')$~\cite{pra75p053808}. We furthermore consider a stationary light source whose statistical fluctuations do not change in time~\cite{Loudon}. The correlation function therefore only depends on the time difference $\tau=t'-t$ of the measurement between $b_0$ and $b_1$ and the time argument $t$ can be neglected.  In the measurements, a single photon counting detector records the number of photons within a time interval $\Delta t$. Both $\langle \hat n(\Delta t) \rangle$ as well as the variance in the photon number fluctuations, $\Delta n^2(\Delta t)=\langle \hat n^2(\Delta t) \rangle-\langle \hat n(\Delta t) \rangle^2$, are obtained by counting photons in many time intervals of length $\Delta t$. The ratio, $F(\Delta t)\equiv\Delta n^2(\Delta t)/\langle \hat n(\Delta t)\rangle$, is called the Fano factor and gauges the magnitude of the photon fluctuations. $F(\Delta t)\ge1$ corresponds to light sources with super-Poissonian photon statistics that are of concern here, while $F(\Delta t)<1$ is found for purely nonclassical light sources exhibiting sub-Poissonian photon statistics.

Multiple scattering of super-Poissonian light is predicted to lead to photon correlations between different angular directions of the output beams.  Having an incident light wave in direction $a$, we derive for $\tau=0$~\cite{smolka}
\begin{equation}\label{eq02}
\overline{C^{Q}_{b_0b_1}(0,\Delta t)}=\left[1+\frac{F(\Delta t)-1}{\langle \hat n(\Delta t)\rangle}\right]\times \overline{C_{b_0b_1}^{(C)}}.
\end{equation}
The first term on the right side is related to the photon statistics of the light source and is the equivalent to the second order coherence function, $g^{(2)}(\tau=0)$~\cite{Loudon}, while the latter factor,
\begin{equation}
\overline{C_{bb'}^{C}}=\frac{\overline{\left<\hat n_{b}(\Delta t)\right>\left<\hat n_{b'}(\Delta t)\right>}}{\overline{\left<\hat n_{b}(\Delta t)\right>}\times\overline{\left<\hat n_{b'}(\Delta t)\right>}},
\end{equation}
is due to classical intensity correlations~\cite{revmodphys71p313}. In the diffusive regime classical intensity correlations vanish and $\overline{C_{bb'}^{C}}=1$. For the special case of a thermal light source Eq.~\eqref{eq02} simplifies to~\cite{Loudon}
\begin{equation}
\overline{C_{b_0b_1}^{Q}(\tau,\Delta t)}= \left[1 + \exp\left(-\pi(\tau/\tau_c)^2\right) \right] \times \overline{C_{b_0b_1}^{(C)}},\label{eq03}
\end{equation}
with $\tau_c$ being the coherence time. In the experiment $\Delta t$ can be chosen in such a way that the Fano factor of the light source, $F(\Delta t)$, corresponds to thermal light, i.e. $F(\Delta t)=\langle \hat n(\Delta t)\rangle+1$. In contrast photons of a coherent light source are uncorrelated ($F(\Delta t)=1$) and the angular photon correlation function equals the classical intensity correlation function, cf. Eq.~\eqref{eq02}.

The experimental setup is displayed in Fig.~\ref{fig01}. As a light source we use a continuous wave Ti:Sapphire laser $(\lambda=780\,\text{nm})$ that is focused onto a ground glass plate. Super-Poissonian photon statistics is obtained by superimposing coherent beams with random amplitudes and phases. For that purpose the ground glass plate is slowly rotated and only a fraction of the transmitted light is collected using an iris with an aperture that is smaller than the average speckle size of the scattered light generated by the ground glass plate~\cite{prl15p912}. The iris ensures that only the part of the light is collected for which wave vectors exhibit maximum momentum correlation~\cite{physreva70p051802}. The light beam is then collimated and directed onto the front surface of a multiple scattering medium consisting of titanium dioxide nano-particles (thickness, $L=6.3\pm0.2\,\mu\text{m}$, transport mean free path, $\ell=0.9\pm0.1\,\mu\text{m}$~\cite{optexpress16p1222}). Two single photon counting detectors (D$_1$, D$_2$) are positioned behind the sample to record photon coincidence counts between multiply scattered photons at different angles. Detector D$_1$ can be rotated to vary the angle $\Delta\Theta$ between both detectors. We avoid the contribution from ballistic propagation of light through the multiple scattering medium by only collecting light polarized perpendicular to the incident light polarization.

\begin{figure}[]
  \centering
  \includegraphics[width=0.45\textwidth]{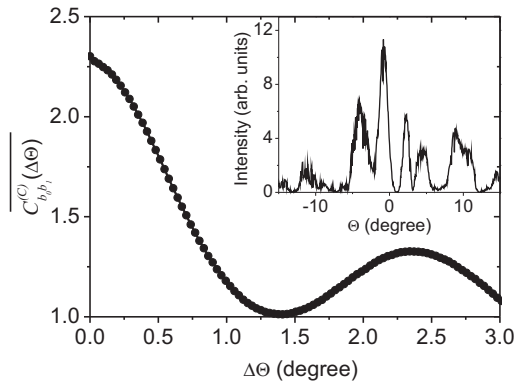}
\caption{Classical intensity correlation function, $\overline{C^{(C)}_{b_0b_1}(\Delta\Theta)}$ versus $\Delta \Theta$ for the speckle pattern shown in the inset. The full width at half maximum represents the average width of the speckles, $\alpha$. The speckle pattern is obtained by rotating detector D$_1$ behind the sample. $\Theta=0^\circ$ corresponds to the configuration where D$_1$ is positioned on the optical axis of the incident light source. The uncertainty in $\overline{C^{(C)}_{b_0b_1}(\Delta\Theta)}$ does not depend on the angle and yields $\pm 0.4$.}
\label{fig02}
\end{figure}

We ensure to probe different speckles, i.e. different output modes $b_0$ and $b_1$, by determining the average speckle width, $\alpha$. A typical intensity speckle pattern of multiply scattered light is shown in the inset of Fig.~\ref{fig02}, recorded by rotating the detector D$_1$. From the angular-resolved speckle pattern we obtain the classical intensity correlation function, $\overline{C^{(C)}_{b_0b_1}(\Delta \Theta)}$~\cite{prl61p834,physreve49p4530} by calculating the auto correlation function of the speckle pattern versus angle, $\Delta\Theta$, plotted in Fig.~\ref{fig02}. The full width at half maximum of the classical intensity correlation function corresponds to the average speckle width, which is $\alpha=1.5^\circ$. The correlation function approaches $\overline{C^{(C)}_{b_0b_1}(\Delta \Theta)}=1$ for $\Delta\Theta\approx1.25$ showing that different speckles are uncorrelated when considering only intensity measurements as opposed to the correlation measurements reported below. The increase in the correlation function for large angles originates from the limited number of investigated speckles and could be suppressed by using better angular filtering with a smaller iris in front of the detector or by probing a larger angular range. The apertures in front of the photon counting detectors D$_1$ and D$_2$ are adjusted to be smaller than $\alpha$ while in the following experiments, the angle between both detectors, $\Delta \Theta$, is chosen to be much larger than $\alpha$. Thereby independent output modes are probed and the classical intensity correlations within a single speckle (see Fig.~\ref{fig02}) do not contribute to the angular photon correlation function.

In order to characterize the properties of the light source, we measure the photon statistics, i.e., we create a histogram from the number of photons, $\hat n(\Delta t)$, that are detected within $\Delta t$. To this end, the sample is removed and D$_1$ is positioned in the light beam emerging from the source after attenuation with a neutral density filter. Figure~\ref{fig01}b displays the measured super-Poissonian photon statistics for $\Delta t=266\,\mu$s yielding $F_\text{OD}(\Delta t)=\langle \hat n_\text{OD}(\Delta t)\rangle=29$. The corresponding Fano factor of the light source without attenuation is calculated using the separately measured optical density OD$=2\times10^{-5}$ of the filter and we obtain  $F(\Delta t)\equiv(F_\text{OD}(\Delta t)-1)/\text{OD}+1=1.4\times 10^6$ equal to the mean number of photons, $\langle \hat n(\Delta t)\rangle=\langle \hat n_\text{OD}(\Delta t)\rangle/\text{OD}$. Hence, the light source incident on the multiple scattering medium exhibits the properties of thermal light and is called a Gaussian radiation source~\cite{prl15p912}. By varying $\Delta t$, the photon counting distribution changes. Figure~\ref{fig01}c shows for example the measured (attenuated) photon statistics for $\Delta t=100\mu$s with a corresponding Fano factor of $F_\text{OD}(\Delta t)=17$ and mean number of photons of $\langle \hat n_\text{OD}(\Delta t)\rangle=10$. By removing the ground glass plate in the incident light beam, we measure $F_\text{OD}(\Delta t)=1$, independent of $\Delta t$, reflecting the coherent state of the continuous wave Ti:Sapphire laser.

The correlation between photons that exit the multiple scattering medium under a relative angular difference of $\Delta \Theta$ is obtained by recording simultaneously the mean number of photons within $\Delta t$ in each direction, $\overline{\langle\hat n_b(\Delta t)\rangle}$. In addition we measure the photon coincidence counts, $\overline{\langle\hat n_{b_0}(\Delta t)\hat n_{b_1}(\Delta t)\rangle}$, within the time-interval $\Delta t$ between detector D$_1$ and D$_2$, see Eq.~\eqref{eq01}. The ensemble average is achieved after repeating the measurement at $200$ different sample positions (indicated by the arrows at the sample in Fig.~\ref{fig01}) corresponding to different realizations of disorder. The experimental configuration further allows us to measure the classical intensity correlations, $\overline{C^{(C)}_{b_0b_1}}$, independently, as explained above. These measurements are fundamentally different from the photon coincidence measurements sine we record the individual ensemble-averaged mean number of photons $\overline{\left<\hat n_{b}(\Delta t)\right>}$ on each detector and the joint ensemble average $\overline{\langle\hat n_{b_0}(\Delta t)\rangle\langle\hat n_{b_1}(\Delta t)\rangle}$ on both detectors.

\begin{figure}[]
  \centering
  \includegraphics[width=0.45\textwidth]{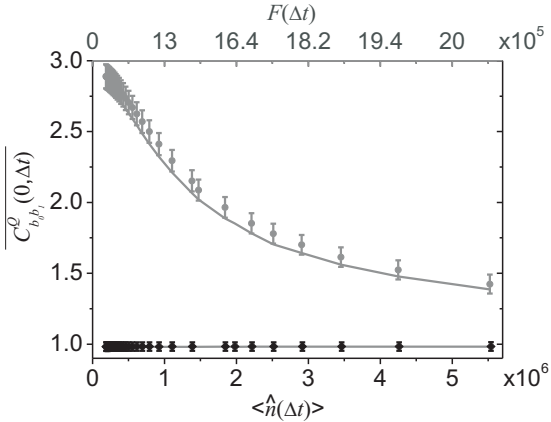}
\caption{Photon correlation function $\overline{C^{Q}_{b_0b_1}(0,\Delta t)}$ (gray circles) depending on $\left<\hat n(\Delta t)\right>$ and $F(\Delta t)$ (top axis) of the light source that is obtained by varying the time interval $\Delta t$ while slowly rotating the ground glass plate, see Fig.~\ref{fig01}a. For a coherent state $\overline{C^{Q}_{b_0b_1}(0,\Delta t)}$ is independent of $\left<\hat n(\Delta t)\right>$ (black squares). The angle between the detectors is chosen to be $\Delta\Theta=9^\circ$ and the theoretical predictions contain no free fitting parameters (solid curves).}
\label{fig03}
\end{figure}

First, we study the angular photon correlation function, $\overline{C^{Q}_{b_0b_1}(0,\Delta t)}$, depending on the average number of photons and Fano factor of the incident light field by varying $\Delta t$, as shown in Fig.~\ref{fig03}. For light sources with super-Poissonian probability distributions $(F(\Delta t)>1)$ we measure a decrease in the strength of the photon correlation function with increasing number of photons and Fano factor (gray data points). On the contrary, we observe that for a coherent light source, $\overline{C^{Q}_{b_0b_1}(0,\Delta t)}$ is independent on the incident number of photons. In order to compare the experimental data with theory we measure also the classical intensity correlations induced by the multiple scattering medium. Afterwards, the Fano factor of the light source is determined with detector D$_1$ as a function of $\Delta t$  by removing the sample. The experimental data are found to be in good agreement with Eq.~\eqref{eq02} without any adjustable fitting parameters. The slightly smaller predicted values of the angular photon correlations (gray line in Fig.~\ref{fig03}) are attributed to variations in the rotation speed of the ground glass plate over time, which influence the Fano factor. For a very low number of incident photons we observe a saturation at $\overline{C^{Q}_{b_0b_1}(0,\Delta t)}\approx2.9$ corresponding to very short time intervals $\Delta t$. This effect we attribute to the properties of the light source determining its photon statistics. The phase of an otherwise coherent light source is scrambled by the rotating glass plate imposing an upper limit on the ratio $(F(\Delta t)-1)/\langle \hat n(\Delta t)\rangle$ and the correlations, respectively, for times $\Delta t$ smaller than the time scale on which the phase distortion occurs. These results demonstrate that the photon fluctuations of multiple scattered light depend sensitively on the investigated timescales, even though samples with static disorder are probed. Thus, the strength of the angular photon correlation function can be controlled by varying the bin time $\Delta t$ in the measurements.

\begin{figure}[]
  \centering
 \includegraphics[width=0.45\textwidth]{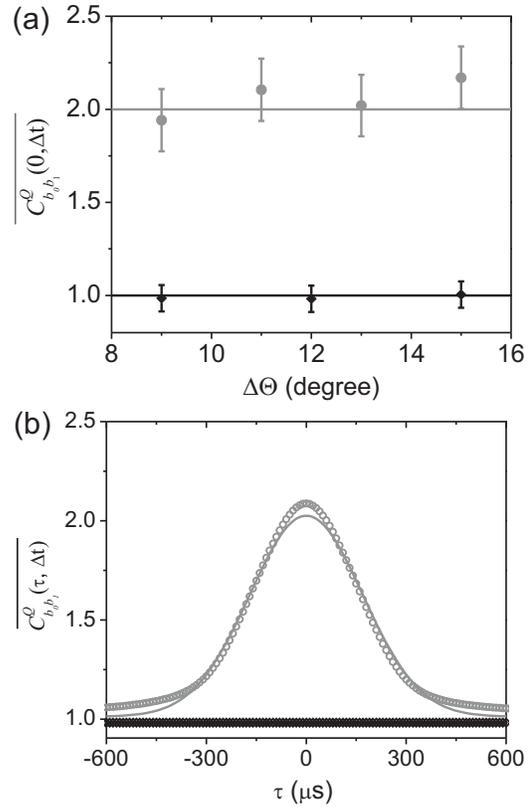}
\caption{\textbf{a}, Angular dependence of $\overline{C^{Q}_{b_0b_1}(0,\Delta t)}$ for a coherent light source (black squares) and a Gaussian radiation source (gray circles). Detector D$_1$ is rotated behind the sample to change $\Delta\Theta$. The straight lines are the theoretical predictions assuming that the classical intensity correlations equal unity. \textbf{b}, Measured temporal dependence of the angular photon correlation function for a Gaussian radiation source (gray circles) and a coherent light source (black squares). Each data point represents an individual  measurement using $\Delta\Theta=9^\circ$ and $\Delta t=266\,\mu$s. For visibility the uncertainties of the data points are not plotted which are $\pm0.03$ for the coherent light source and $\pm0.07$ for the Gaussian radiation source. The gray curve is a fit to the data with the coherence time as a free parameter.}
\label{fig04}
\end{figure}

Figure~\ref{fig04}a plots the main result of this report, i.e. the angular photon correlations as a function of the angle $\Delta\Theta$ between detector D$_1$ and D$_2$ using a coherent and a Gaussian radiation source with $\Delta t=266\,\mu$s. Here detector D$_1$ is rotated behind the sample, as shown in Fig.~\ref{fig01}a. For each detector position we recorded for 200 different realizations of disorder the individual as well as joint photon counting distributions (see Fig.~\ref{fig01}b). We neither observe for the coherent light source nor for the Gaussian radiation source a dependence of the angular photon correlation function on the angle. This holds in the diffusive regime of multiple scattering, where the classical intensity correlation function does not vary with angle, $C^{(C)}_{b_0b_1}=1$. For a light source with super-Poissonian photon statistics the photon correlation function differs from unity demonstrating that different far-field speckles are strongly correlated and the ensemble-averaged correlation does not decrease as the angle increases. We underline that the relative angle, $\Delta\Theta$, between the detectors is much larger than the speckle width, $\alpha$, i.e. we probe different speckles that are classically uncorrelated. These new correlations do not depend on the angle between the speckles and are therefore infinite in range, in agreement with theoretical predictions (straight lines in Fig.~\ref{fig04}a). Thereby, we show that a multiple scattering medium translates the temporal photon correlations of the light source into angular correlations connecting simultaneously all input and output modes to each other - a property that has been suggested as a way to generate quantum entanglement in a disordered medium~\cite{prl105p090501,prl105p163905}. The error bars in $\overline{C^{Q}_{b_0b_1}(0,\Delta t)}$ arise mainly from uncertainties in the classical intensity correlations caused by a finite number of disorder realizations for the ensemble average. In previous work on with squeezed light the photon correlation function was extracted from total transmission quantum noise measurements~\cite{prl102p193901}, and not through direct photon counting as employed here. We note that the technique of measuring  the quantum noise of the total transmission to extract the photon correlation function only applies in a diffusive regime, while the direct detection scheme developed here can be applied also in mesoscopic and localized regimes where interesting new quantum correlations are predicted~\cite{prl95p173901}. Furthermore, direct photon counting measurements would be required for testing recent predictions that quantum interference can survive multiple scattering even after ensemble averaging over configurations of disorder~\cite{prl105p090501}.

Our experiments on the photon correlations after multiple scattering are potentially relevant to the field of ghost imaging, where the second order coherence between two separate, but spatially correlated light beams, can be used to spatially resolve an object in one of the beams~\cite{prl94p063601}. A light beam emerging from a pseudothermal light source, similar to the one used here but without an iris behind the ground glass plate (see Fig.~\ref{fig01}a), is divided in two optical paths using a beam splitter~\cite{physreva70p051802}. In one of the paths the object is placed in front of a large bucket detector and the second light beam is imaged with a lens onto a photo detector with a small aperture. Since this light source is chaotic, only photons within a single speckle spot of the light exiting the ground glass plate are spatially correlated, resulting in point-to-point correspondence of the photon correlations between object plane and image plane~\cite{prl94p063601}. By scanning the detector across the image plane, the object can be resolved from the spatial dependence of the photon correlations. Our experiments are distinct to this scheme since the photon correlation function observed here extends over the entire speckle pattern that is generated by multiple scattering, and the implemented iris ensures that only photons of the light source with similar wave vectors are selected. The infinite range of the angular photon correlation thus originates from the photon statistics of the incident light beam distributed over the available degrees of freedom due to the multiple scattering~\cite{prl95p173901,physreports336p1}. Thus, if multiple scattering occurs after the pseudothermal light source, the point-to-point mapping of the second order coherence is lost and cannot be exploited for ghost imaging. A second approach to ghost imaging does not utilize photon fluctuations of the light source but uses a multiple scattering medium in front of the beam splitter. In this case the generated intensity speckle pattern is collimated and split in two light beams, establishing now a point-to-point correspondence of the classical light intensity between object and image plane. After performing an ensemble average over many different speckle patterns, the resultant classical spatial intensity correlation function resolves the object~\cite{prl94p183602}. These two fundamentally different ghost imaging approaches utilize the photon correlations of the light source and intensity correlations induced by multiple scattering, respectively. Here, we use a pseudothermal light source to probe angular photon correlations that are induced by a multiples scattering medium and study its dependence on the photon fluctuations of the incident quantum state of light, cf. Fig.~\ref{fig03}. While ghost imaging schemes so far exploit correlations between two light beams, multiple scattering would in general allow to induce correlations between $N$ output modes obtained by mixing N independent quantum states.

In order to investigate the temporal dependence of the angular photon correlations induced by a multiple scattering medium, a time delay, $\tau$, is induced between detector D$_1$ and D$_2$. Figure~\ref{fig04}b presents the angular photon correlation function depending on $\tau$ using a Gaussian radiation source (gray data points). As the time difference increases, a clear decay of $\overline{C^{Q}_{b_0b_1}(\tau,\Delta t)}$ is observed. Fitting the experimental results with theory (Eq.~\eqref{eq03}), we determine the coherence time to be $\tau_c=750\:\mu\text{s}$.
The coherence time describes the dephasing of a light source, i.e. the average time interval between phase distortion fluctuations. Only coherence times of the light source can be resolved that are larger than $\Delta t$, i.e., the experimentally measured photon statistic does not contain information about photon fluctuations on shorter timescales. As $\tau_c>\Delta t$ we resolve the coherence time of our Gaussian radiation source. The time response of $\overline{C^{Q}_{b_0b_1}(\tau,\Delta t)}$ originates from the temporal bunching of photons in the thermal state. For comparison, the temporal dependence of the angular photon correlation function is investigated for a coherent light source, too (black data points, Fig.~\ref{fig04}b), where as expected the angular photon correlation function is independent of $\tau$ and equals unity.

In conclusion, we demonstrated experimentally the first direct measurement of angular photon correlations that are induced by multiple scattering of light. The time correlations of the multiply scattered photons at different angles were found to be dependent on the photon statistics of the incident light source. In excellent agreement with the quantum theory of multiple scattering, we observed that the photon correlations are infinite in range in the diffusive regime. Combining the quantum aspects of multiple scattering with phase shaping techniques~\cite{naturephoton4p320} could open a new route to vary the quantum state of the multiple scattered light. Our angular-resolved experimental technique is crucial to study experimentally predicted quantum interference and quantum entanglement in three-dimensional multiple scattering media~\cite{prl105p163905,prl105p090501,arXiv10072085v2}.

The authors thank Johan R. Ott, A. Huck, and P.S. Scalia for stimulating discussions, and gratefully acknowledge the Danish Council for Independent Research (Technology and Production Sciences and Natural Sciences) for financial support.


\begin{thebibliography}{99}

\bibitem{physreports238p135} R. Berkovits and S. Feng, Phys. Reports {\bf 238}, 135 (1994).
\bibitem{revmodphys71p313} M.C.W. van Rossum and T.M. Nieuwenhuizen, Rev. Mod. Phys. {\bf 71}, 313 (1999).
\bibitem{prl61p834} S. Feng, C. Kane, P. A. Lee, and A. D. Stone, Phys. Rev. Lett. {\bf
 61}, 834 (1988).
 \bibitem{prl81p1829} C.W.J. Beenakker, Phys. Rev. Lett. {\bf 81}, 1829 (1998).
\bibitem{prl95p173901} P. Lodahl, A.P. Mosk, and A. Lagendijk, Phys. Rev. Lett. {\bf 95}, 173901 (2005).
\bibitem{prl102p193601} C.W.J. Beenakker, J.W.F. Venderbos, and M. P. van Exter, Phys. Rev. Lett. {\bf 102}, 193601 (2009).
\bibitem{prl105p163905} Y. Lahini, Y. Bromberg, D.N. Christodoulides, and Y. Silberberg, Phys. Rev. Lett. \textbf{105}, 163905 (2010).
\bibitem{arXiv10072085v2} N. Cherroret and A. Buchleitner, arXiv:1007.2085v3 (2010)
\bibitem{prl105p090501} J.R. Ott, N.A. Mortensen, and P. Lodahl, Phys. Rev. Lett. \textbf{105}, 090501 (2010).
\bibitem{science287p287} A.L. Moustakas, \emph{et al.}, Science 287, 287 (2000).
\bibitem{prl89p043902} J. Tworzydlo and C. W. J. Beenakker, Phys. Rev. Lett. {\bf 89}, 043902 (2002).
\bibitem{prl102p193901} S. Smolka, A. Huck, U.L. Andersen, A. Lagendijk, and P. Lodahl, Phys. Rev. Lett. {\bf 102}, 193901 (2009).
\bibitem{prl94p153905} P. Lodahl and A. Lagendijk, Phys. Rev. Lett. {\bf 94}, 153905 (2005). %
\bibitem{prl97p103901} S. Balog, P. Zakharov, F. Scheffold, and S.E. Skipetrov, Phys. Rev. Lett. {\bf 97}, 103901 (2006).
\bibitem{science327p1352} L. Sapienza, \emph{et al.}, Science {\bf 327}, 1352 (2010).
\bibitem{prl104p173601} W.H. Peeters, J.J.D. Moerman, and M.P. van Exter, Phys. Rev. Lett. {\bf 104}, 173601 (2010). 
\bibitem{nature177p27} R. Hanbury Brown and R.Q. Twiss, Nature {\bf 177}, 27 (1956).
\bibitem{physreports336p1} Ya.M. Blanter and M. B\"uttiker, Physics Reports \textbf{336}, 1 (2000).
\bibitem{physreva52p3429} T.B. Pittman, Y.H. Shih, D.V. Strekalov, and A.V. Sergienko, Phys. Rev. A \textbf{52}, R3429 (1995).
\bibitem{prl94p063601} A. Valencia, G. Scarcelli, M. D\'Angelo, and Y. Shih, Phys. Rev. Lett., \textbf{94}, 063601 (2005).
\bibitem{prl94p183602} F. Ferri, D. Magatti, A. Gatti, M. Bache, E. Brambilla, and L.A. Lugiato Phys. Rev. Lett., \textbf{94}, 183602 (2005).
\bibitem{scully} M.O. Scully and M.S. Zubairy, \emph{Quantum Optics} (Cambridge University Press, 1997).
\bibitem{pra75p053808} S.E. Skipetrov, Phys. Rev. A \textbf{75}, 053808 (2007).
\bibitem{Loudon} R. Loudon, \emph{The Quantum Theory of Light} (Oxford University Press, 2000).
\bibitem{smolka} S. Smolka \emph{et al.} (to be published).
\bibitem{prl15p912} F.T. Arecchi, Phys. Rev. Lett. {\bf 15}, 912 (1965).
\bibitem{physreva70p051802} G. Scarcelli, A. Valencia, and Y. Shih, Phys Rev. A, \textbf{70}, 051802 (2004).
\bibitem{optexpress16p1222} O.L. Muskens and A. Lagendijk, Optics Express,  {\bf 16}, 1222 (2008).
\bibitem{physreve49p4530} J.H. Li and A.Z. Genack, Phys. Rev. E {\bf 49}, 4530 (1994).
\bibitem{naturephoton4p320} I. M. Vellekoop, A. Lagendijk, and A. P. Mosk, Nature Photonics \textbf{4}, 320 (2010).

\end{thebibliography}
\end{document}